\documentclass[12pt]{article}
\usepackage{epsfig}
\newcommand{\be}{$\beta{}$}

\newcommand{\fm}{f\!m}
\newcommand{\qs}{$\hat{q}(x)$}
\newcommand{\qsa}{$\left|\hat{q}(x)\right|$}

\begin{document}
\begin{center}
\Large
Gauge Singularities in the SU(2) Vacuum on the Lattice\\
\vspace{0.6cm}
\large
 F. Gutbrod\\
 Deutsches Elektronen-Synchrotron DESY\\
\small
 Notkestr. 85, D22603 Hamburg, Germany\\
 e-mail: gutbrod@mail.desy.de\\
\vspace{1.cm}
\end{center}\hspace{1.cm}
\parbox{18.cm}

\abstract{I summarize and extend
          the qualitative results, obtained previously by inspection
          of SU(2) lattice gauge field configurations.
          These configurations were generated by the Wilson action,
          then transformed to a Landau gauge and smoothed by Fourier
          filtering. This leads to sharp peaks in field strengths
          and related quantities, the characteristics of which are very
          well separated from a background. These spikes are caused
          by gauge singularities, Their densiis determined
          as $1.5/f\!m^4$, with very good scaling properties
          as a function of the bare coupling constant.
          The number of spikes within a configuration vanishes
          when approaching the deconfinement region.
          Furthermore, the Landau-gauging procedure becomes unique,
          if the probability to find a spike is much smaller than unity.
          The relation of the spikes to the instantons
          obtained by cooling is investigated.
          Finally, a correlation between the presence of spikes
          and the infrared behaviour of the gluon propagator
          is demonstrated.}
\normalsize
\section{Introduction}
\label{intro}
\begin{figure}
\begin{center}
\epsfig{file=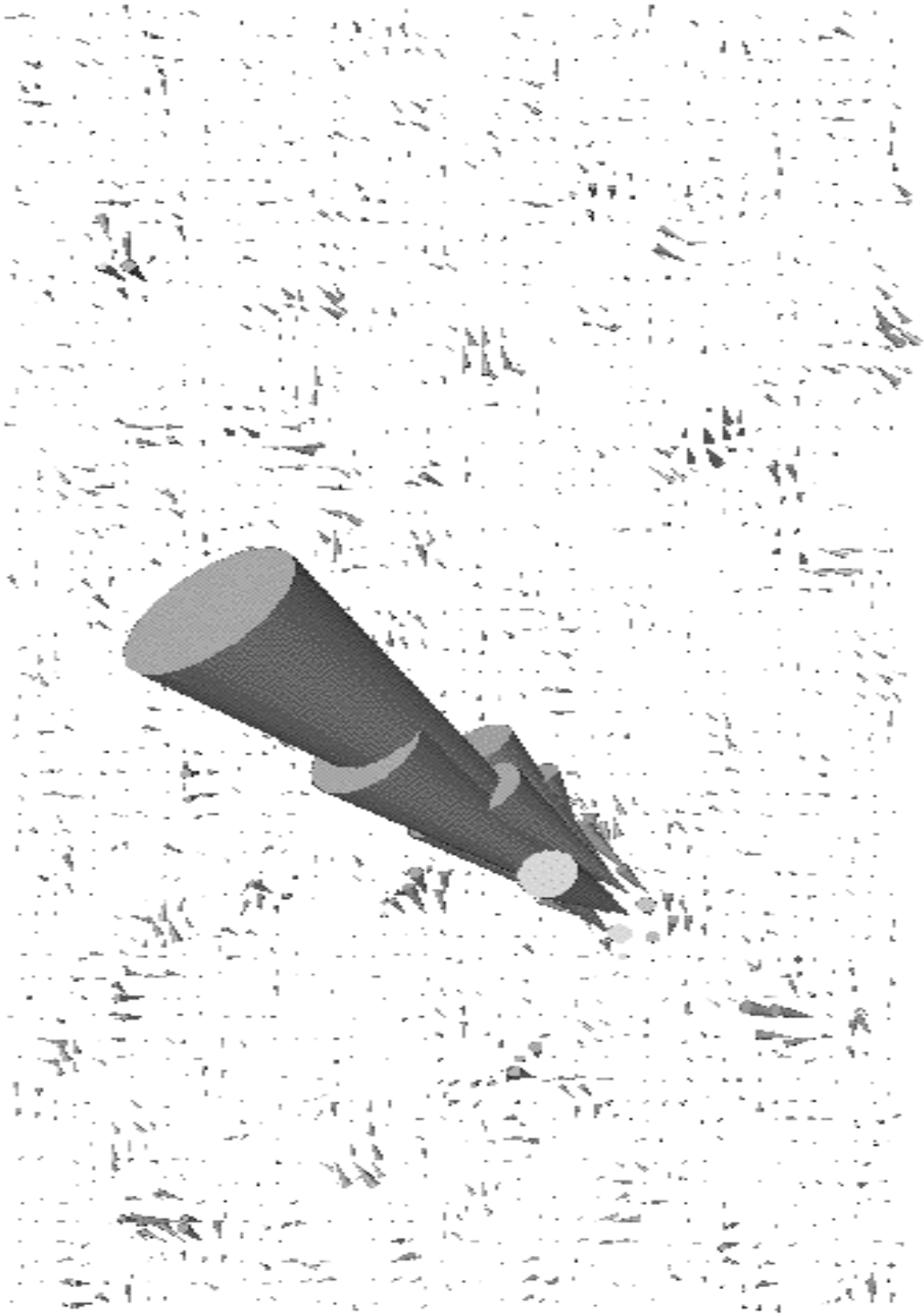,height=14cm,width=20cm,angle=90}
\parbox{14cm}{\caption{\label{spike}
The topological charge density
on a lattice plane running through a typical spike. The direction of the
cones indicates the orientation of the charge in
three-dimensional colour space.
The lattice size is $48^3 \times 64$, and $\beta$ = 2.85.
The cut-off for the Fourier transformed gauge fields is
$\lambda^2 = 0.5 a^2$.}}
\end{center}
\end{figure}

If one transforms a single configuration of SU(2) lattice
gauge fields to a Landau gauge and then removes the high
momentum Fourier components of the gauge fields by some exponential
cut-off ("Fourier filtering", F.f.,  or "smoothing"),
one observes the following phenomenon \cite{GUTBROD1}:
For sufficiently large lattices and/or for sufficiently large
bare coupling constants, at a couple of lattice positions
the gauge fields show a rapid
variation in all space-time directions and for all
colour components. These variations lead to narrow spikes
in the Wilson action density $S(x)$ and
in the topological charge density\footnote{The normalization
of these two quantities will be given in
eqs. \ref{action} and \ref{topol}.} \qs{}.
As an example, in fig. \ref{spike} the charge density
\qs{} is shown for a plane of a large lattice.
The cones have been
generated by the VRML-language\cite{AMES} and visualized by the
browser GLVIEW\cite{GRAHN}.
The positions of almost all the spikes
do not vary in the different gauges
which are due to different Gribov copies \cite{GRIBOV}.
For sufficiently modest filtering, the spatial
extension of these spikes amounts to one or two lattice
units, which roughly corresponds to $0.05 f\!m$.

Due to Fourier smoothing, the quantities
$S(x)$ and \qs{} are gauge dependent, and their physical
interpretation is not straightforward.
The observation that, within the spikes, one observes self-duality, i.e.
\qs{} $ \approx \pm S(x)$ within a 10$\%$ accuracy, might suggest
that we observe narrow instantons.
Two facts contradict this interpretation. First,
one observes that the distribution in terms of $S(x)$
-summed over a few lattice sites around the spikes-
is strongly peaked for large action values. Those correspond  to
small-sized instanton-like objects.
For details see section \ref{density}.
On the contrary, several lattice
studies in SU(2) predict a size distribution with
a maximum for sizes in the range from 0.2 $f\!m$
\cite{DEGRAND} to  0.4 $f\!m$ \cite{FORCRAND} and  around 0.3 $f\!m$
or higher in SU(3) \cite{AHAS}. Secondly, if the configuration
is smoothed by the standard cooling technique \cite{TEPER},
then one indeed observes instantons at the position of the spikes,
but their size is much larger than that of the spikes.

If one transforms to a Landau gauge prior to cooling,
one can directly observe that the gauge fields
are singular (regulated by the lattice) at the positions of the spikes.
All observations point towards the interpretation that
the Landau gauging leads to singularities in the gauge fields,
if the lattice parameters are within the confinement region.
A gauge singularity at the origin
can be created by imposing the gauge transformation\footnote{The origin
$ x = 0$ has to be chosen apart from a lattice site. $\vec{\tau}$
are the Pauli matrices.}
\begin{equation}
\label{singau}
g(x) = (x_0 + i\vec{x}\vec{\tau})/\left|x\right|
\end{equation}
on a relatively smooth gauge field configuration,
e.g. on a large-size instanton in the regular gauge\footnote{The
generalization to several singularities at different positions is obvious.}.
Then, close to the origin, there occurs
a cancellation between the derivative
terms in the action and the cubic and higher gauge field terms.

This cancellation will be mutilated by filtering,
which acts differently on the two contributions.
Accordingly, a singular gauge field distribution will
develop, under F.f., a sharp peak in the action and in \qs,
even if both quantities are smooth in $x$ without filtering\footnote{This
phenomenon has been checked by analytic calculations in the continuum
by H. Joos.}.
Thus, the plaquette action and \qs{} -after smoothing- no longer possess
a definite physical meaning. They have to be understood
as a measure for unexpected things that happen while gauging.

In this paper, I will concentrate on the phenomenology of the spikes,
i.e. on their density and on their correlation with other nonperturbative
phenomena. The following topics will be treated:
\begin{enumerate}
\item
For large \be = $4/g^2_0$, the spikes are very well defined objects,
if the action (or the topological charge density)
-summed over a few lattice units around
the action maximum- is taken as a probe.
This is demonstrated by the histogram in fig.~\ref{histogram},
which is based on $O(30)$ configurations
on large lattices and large~\be.
The histogram displays a well defined valley
between a peak containing the
spikes (all satisfying self-duality \qs$ \approx \pm S(x)$),
and a background which contains distinctly different parts
of the lattice configuration. This allows
to determine the density of spikes as
\begin{equation}
\label{rho}
\rho_{spikes}= (1.5 \pm 0.20) / f\!m^4.
\end{equation}
Since the average number of spikes per configuration
is well determined, i.e. without a significant
systematic error, the density of spikes
in physical units can be used for a scaling study
as a function of \be. Very good scaling with
the string tension results \cite{BOOTH}
is observed in the interval
$ 2.70 \leq  \beta \leq 2.85 $.
For details see section \ref{density}.
The relation of the
number of spikes and the deconfinement transition
will also be discussed in that section.

\item The density of spikes is correlated with the number
of Gribov copies which are found during the Landau gauging
process. If the lattice is chosen so small that the probability
for finding one or more spikes
in the configuration is much less than unity,
then, empirically, the gauging procedure is practically unique.
This means that the Landau gauging algorithm
ends up -with high probability- preferentially in the same gauge,
i.e. one rarely observes the appearance of Gribov copies \cite{GRIBOV}
if only a finite number of gauging trials is performed
(a trial is defined by applying a random gauge
prior to the Landau gauging). The observation is that
if copies are found, the probability to find spikes is enhanced too.
This positive correlation between the appearance of spikes and copies
will be demonstrated in section \ref{gribov}.
\item
It is instructive to study alternatives
to the Fourier filtering technique,
notably the cooling technique \cite{TEPER}.
In section \ref{cooling}, I will show that there is an interesting
connection between the outcome of cooling and of F.f.
If one starts from a configuration in the Landau gauge, finds
the position of its spikes and then cools the unfiltered lattice
configuration, one observes the following: The absolute values of
the gauge fields, $\left|\vec{u}_{\mu}(x)\right|$ (see eq. \ref{pauli}),
decrease under cooling almost everywhere,
except at the positions of the spikes. There,
gauge singularities show up.
The local action maxima, which always accompagny the singularities,
are broad in most cases.

\item The fact that the spikes manifest themselves as zeros
in the gauge fields -in all directions and for all colours-
suggests that the gluon propagator and other
correlators of the gauge fields
behave differently if spikes are present
in a configuration or if not.
This is the case, as it will be shown in section~\ref{correlations}.
\end{enumerate}\section{Landau Gauge and Filtering}
Here a couple of topics relevant for gauging and filtering
will be discussed.
\label{gauging}
The Landau gauging algorithm on the lattice maximizes the sum of
the "large" SU(2)-components $u_{0,\mu}(x)$
in the link representation
\begin{equation}
U_{\mu}(x)  = u_{0,\mu}(x) + i\vec{\tau}\vec{u}_{\mu}(x)
\label{pauli}
\end{equation}
with
\begin{equation}
 u_{0,\mu}^2(x) + \vec{u}_{\mu}^2(x) = 1.
\label{norm}
\end{equation}
Thus, one searches numerically for maxima of
\begin{equation}
F(U) = \sum_{x,\mu} u_{0,\mu}(x).
\label{gauge}
\end{equation}
under local gauge transformations $U$.
If the maximum condition is fulfilled, one has
\begin{equation}
\sum_{\mu} \vec{u}_{\mu}(x) - \vec{u}_{\mu}(x - e_{\mu}) = 0,
\label{lattdiver}
\end{equation}
which corresponds to
\begin{equation}
\label{contgauge}
\partial_{\mu}\vec{A}_{\mu} = 0
\end{equation}
in the continuum. At the center of an instanton in the singular gauge,
eq. \ref{contgauge} makes no sense. This fact requires to describe
an instanton around the center by gauge fields in the regular gauge
and at infinity by the singular gauge. On the lattice, however, the
singular gauge is tolerable, if the center is
not located at a lattice site,
and lattice configurations in the Landau gauge seem to choose
this option.

The Gribov ambiguity \cite{GRIBOV} implies, that there exist
-for sufficiently large lattices in physical
units\footnote{In section \ref{gribov} we will see, that
the critical size is about $(0.9\fm)^4$.}- many different
maxima where eq. \ref{lattdiver} is satisfied.
A priori, it is not clear whether gauge dependent average
values of link variables, e.g. the gauge field propagator,
depend strongly on the special Landau gauges,
which are characterized by the different values
of the gauge function eq. \ref{gauge} for a given configuration.
Empirically, the gauge field propagator
does not depend strongly on the special gauge.
This can be demonstrated by studying the correlation
between numerical results for the gluon propagator
and the values of the gauge function $F(U)$ for different
gauges. The latter can be generated either by different gauging algorithms
(e.g. overrelaxation or conjugate gradient) or by different random gauges,
applied before the iterative gauging starts. The following numerical
results have been obtained:

For a lattice of size $24^3\times 32$ at \be = 2.5, corresponding to a
physical size\\ $(1.9)^3~\times~2.5~\fm^4 $, 20 random gauges were
used for each configuration, and the values for the gluon
propagator at the smallest momenta were ordered with respect
to the values of the gauge function, i.e. whether $F(U)$
was smaller or larger than the average value
(which is determined for the individual configuration).
No significant difference between the two data sets could be observed,
with an overall relative accuracy of $5 \%$ for the gluon propagator
at $q^2 = 0$, and with $1.5 \%$ at $q^2 > 0$.
The upper limit for the difference between one of the ordered data
sets and the average over the random gauges is $2 \%$ for $q^2 > 0$.
The stochastic noise, due to measuring on different configurations,
is comparable to that due to measuring at different gauges,
with no significant correlation between the value
of the gauge function and the value of the propagator.

The Landau gauging drives the configuration to the
region of small vector components
$\vec{u}_{\mu}(x)$ in the sense
that link variables with
$u_{0,\mu}(x) < 0 $ become very rare.
For presently available coupling
constants, $ \beta \leq 2.85$,
this goal has almost been achieved.
This allows for a linearization and subsequent
Fourier transformation of the gauge field variables.
Besides the trivial linearization,
\begin{equation}
\vec{u}_{\mu}(x) \Rightarrow \vec{A}_{\mu}(x),
\label{linear}
\end{equation}
I use a stereographic transformation
in order to minimize errors due to residual
negative $u_{0,\mu}(x)$. There
I define gauge fields by
\begin{equation}
\vec{u}_{\mu}  \sqrt{2 / (1 + u_{0,\mu})}
= \vec{w}_{\mu}(x)
\Rightarrow \vec{A}_{\mu}.
\label{stereo}
\end{equation}
This stereographic transformation brings the "south-pole"
$u_0 = - 1$ not to infinity, but to $\vec{w}_{\mu}^2 = 4$.
The Fourier filtering is applied to the gauge fields
$\vec{A}_{\mu}(x) $, (see eq.~\ref{filter}) and after
suppression of high momentum
Fourier amplitudes, the gauge fields  are transformed
back to SU(2)-variables.
After an eventual smoothing of the fields,
$\vec{A}_{\mu} \Rightarrow \vec{A'}_{\mu} $ by F.f.,
normalized link variables will be reconstructed by the obvious inversion:
\begin{eqnarray}
\label{reunit}
\vec{w}_{\mu}  & = & \vec{A'}_{\mu}, \\
\vec{u}'_{\mu} & = & \sqrt{1 - \vec{w}^2_{\mu}/4}\; \vec{w}_{\mu}, \\
u'_{0,\mu}     & = & \pm \sqrt{1 - \vec{u'}_{\mu}^2},\;\;
         - \rm sign \;\; for \;\;  \it \vec{w}^2_{\mu} > 2.
\end{eqnarray}
The differences between the two methods of linearization amount to $5\%$
for the field strengths, with no impact on the existence
and properties of the spikes.

With these variables, the action and the topological charge density
will be defined in the following way: First of all,
the field strength tensor
$F_{\mu\nu}^a(x)$ is calculated by averaging the link
products around the 4 plaquettes $\rho,\sigma$
with $\rho,\sigma \neq \mu,\nu$,
which are connected to $x$.
Given $F^{(a)}_{\mu\nu}(x)$,
electric fields $\vec{E}^{(a)}(x)$ and magnetic fields
$\vec{B}^{(a)}(x)$ are defined, and the action is
\begin{equation}
\label{action}
S(x) = \frac{1}{2}\sum_a(\vec{E}^{(a)}(x) \vec{E}^{(a)}(x) +
\vec{B}^{(a)}(x) \vec{B}^{(a)}(x)).
\end{equation}
For convenience, I define
\begin{equation}
\label{topol}
\hat{q}(x) = \sum_a\vec{E}^{(a)}(x) \vec{B}^{(a)}(x)
\end{equation}
and call it topological charge density, inspite
of the nonstandard normalization. For self-dual objects,
one has $S(x) = \pm \hat{q}(x)$.

Due to the slight non-locality of the operator \qs,
it should have a negative correlator \cite{SEILER}
\begin{equation}
\label{negativity}
<\hat{q}(x) \hat{q}(0)>\; \leq \;0
\end{equation}
only for $x^2 > 4a^2$.
This property has been successfully checked for 10
configurations on large lattices.
The negativity breaks down for filtered and for cooled
configurations\footnote{Obviously, the negativity excludes the
dominance of four-dimensional, "sign coherent structures"
in the true vacuum, as has been emphasized by \cite{HORVATH2}.}.
This can be read off, for instance, from fig. 16 of ref. \cite{FORCRAND},
and it also has been verified for the cooled configurations used in
section \ref{cooling}.
\section{Density of Spikes}
\label{density}
Here it will be shown that the occurence of spikes is
a well defined phenomenon in the sense that the spikes
are separated from a background of action maxima
by a deep valley, such that their density can be determined
without a significant systematic error.
This is true, especially, for large values of \be.
For smaller values, the depth of the valley  diminuishes somewhat,
without a severe loss of significance.
Thus, it can be tested quite well whether the density of spikes
scales under a variation of \be{} in accordance with the
string tension, inspite of the limited number of spikes identified
so far\footnote{This is true because the change of scale
enters the density with the $4^{th}$ power.}.

An interesting question is how the average number of spikes
varies with the lattice volume, if \be{} is kept fixed. The
preliminary evidence is that the density is roughly independent
of the volume, and that the parameters of the
critical volume, where one expects just one spike
per configuration, are close to the deconfinement transition.
This will be discussed at the end of this section.

The following quantitative analysis is based on\\
A) 34 configurations on a large lattice with size
     $48^3\times64$ at \be =2.85,
with twisted boundary conditions \cite{KAGA}, and \\
B) 25  configurations on a smaller lattice with size
   $32^3\times48$ at \be =2.70, also with twisted boundary
conditions.

Each of these configurations has been gauged
to a Landau gauge and then Fourier filtered,
with the following substitution for the momentum space amplitudes
\begin{equation}
\label{filter}
\vec{A}_{\mu}(q) \rightarrow \vec{A}'_{\mu}(q)=
\vec{A}_{\mu}(q) \exp(-\lambda^2 q^2).
\end{equation}
For identifying spikes, I use $\lambda^2 = 0.5 a^2$.
This modest filtering already leads to a strong suppression of
the perturbative noise of the $\vec{A}_{\mu}$.
The filtered gauge fields were
transformed back to x-space and subsequently reunitarized (see eq.
\ref{reunit}).

For each of these filtered configurations, I have identified
$O(500)$ local maxima of the action density $S(x)$.
Then, $S(x)$ has been summed
within a radius of $R = 2a $ around the positions $x_m$
of these maxima,
\begin{equation}
\label{sigdef}
\Sigma_m(\lambda^2,R) = \sum_{x}^{(x-x_m)^2\leq R^2} S(x).
\label{sigma}
\end{equation}

For case A, the distribution of the values of $\Sigma_m(\lambda^2,R)$
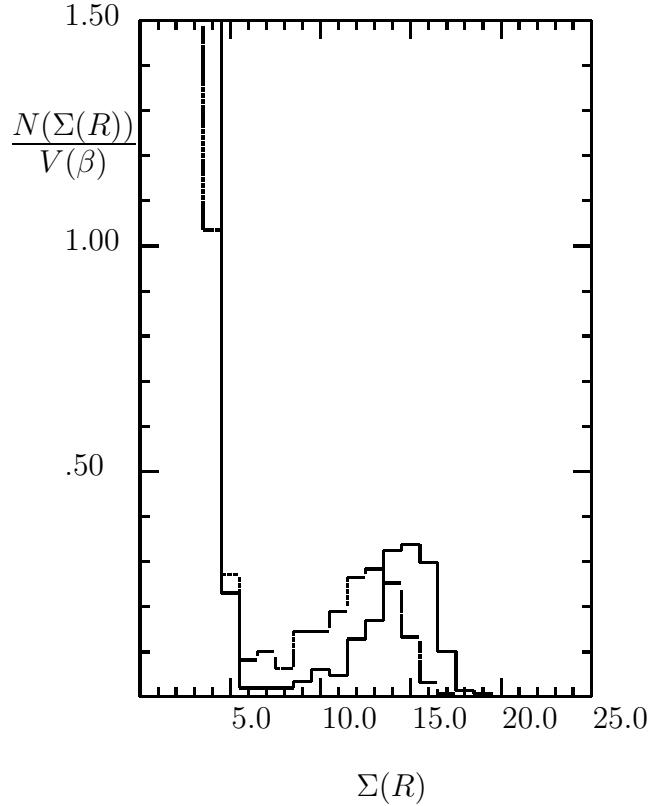
\begin{figure}[t]
\unitlength0.6cm
\begin{center}
\begin{picture}(10.,15.)
\normalsize
\thicklines
\put(  .4000, 0){\line(0, 1){  .2000}}
\put(  .4000,15){\line(0,-1){  .2000}}
\put(  .8000, 0){\line(0, 1){  .2000}}
\put(  .8000,15){\line(0,-1){  .2000}}
\put( 1.2000, 0){\line(0, 1){  .2000}}
\put( 1.2000,15){\line(0,-1){  .2000}}
\put( 1.6000, 0){\line(0, 1){  .2000}}
\put( 1.6000,15){\line(0,-1){  .2000}}
\put( 2.0000, 0){\line(0, 1){  .2000}}
\put( 2.0000,15){\line(0,-1){  .2000}}
\put( 2.4000, 0){\line(0, 1){  .2000}}
\put( 2.4000,15){\line(0,-1){  .2000}}
\put( 2.8000, 0){\line(0, 1){  .2000}}
\put( 2.8000,15){\line(0,-1){  .2000}}
\put( 3.2000, 0){\line(0, 1){  .2000}}
\put( 3.2000,15){\line(0,-1){  .2000}}
\put( 3.6000, 0){\line(0, 1){  .2000}}
\put( 3.6000,15){\line(0,-1){  .2000}}
\put( 4.0000, 0){\line(0, 1){  .2000}}
\put( 4.0000,15){\line(0,-1){  .2000}}
\put( 4.4000, 0){\line(0, 1){  .2000}}
\put( 4.4000,15){\line(0,-1){  .2000}}
\put( 4.8000, 0){\line(0, 1){  .2000}}
\put( 4.8000,15){\line(0,-1){  .2000}}
\put( 5.2000, 0){\line(0, 1){  .2000}}
\put( 5.2000,15){\line(0,-1){  .2000}}
\put( 5.6000, 0){\line(0, 1){  .2000}}
\put( 5.6000,15){\line(0,-1){  .2000}}
\put( 6.0000, 0){\line(0, 1){  .2000}}
\put( 6.0000,15){\line(0,-1){  .2000}}
\put( 6.4000, 0){\line(0, 1){  .2000}}
\put( 6.4000,15){\line(0,-1){  .2000}}
\put( 6.8000, 0){\line(0, 1){  .2000}}
\put( 6.8000,15){\line(0,-1){  .2000}}
\put( 7.2000, 0){\line(0, 1){  .2000}}
\put( 7.2000,15){\line(0,-1){  .2000}}
\put( 7.6000, 0){\line(0, 1){  .2000}}
\put( 7.6000,15){\line(0,-1){  .2000}}
\put( 8.0000, 0){\line(0, 1){  .2000}}
\put( 8.0000,15){\line(0,-1){  .2000}}
\put( 8.4000, 0){\line(0, 1){  .2000}}
\put( 8.4000,15){\line(0,-1){  .2000}}
\put( 8.8000, 0){\line(0, 1){  .2000}}
\put( 8.8000,15){\line(0,-1){  .2000}}
\put( 9.2000, 0){\line(0, 1){  .2000}}
\put( 9.2000,15){\line(0,-1){  .2000}}
\put( 9.6000, 0){\line(0, 1){  .2000}}
\put( 9.6000,15){\line(0,-1){  .2000}}
\put(10.0000, 0){\line(0, 1){  .2000}}
\put(10.0000,15){\line(0,-1){  .2000}}
\put( 2.0000, 0){\line(0, 1){  .4000}}
\put( 2.0000,15){\line(0,-1){  .4000}}
\put( 4.0000, 0){\line(0, 1){  .4000}}
\put( 4.0000,15){\line(0,-1){  .4000}}
\put( 6.0000, 0){\line(0, 1){  .4000}}
\put( 6.0000,15){\line(0,-1){  .4000}}
\put( 8.0000, 0){\line(0, 1){  .4000}}
\put( 8.0000,15){\line(0,-1){  .4000}}
\put(10.0000, 0){\line(0, 1){  .4000}}
\put(10.0000,15){\line(0,-1){  .4000}}
\put( 1.8000, -.7000){    5.0}
\put( 3.8000, -.7000){   10.0}
\put( 5.8000, -.7000){   15.0}
\put( 7.8000, -.7000){   20.0}
\put( 9.8000, -.7000){   25.0}
\put( 0, 1.0000){\line( 1, 0){ .2000}}
\put(10, 1.0000){\line(-1, 0){ .2000}}
\put( 0, 2.0000){\line( 1, 0){ .2000}}
\put(10, 2.0000){\line(-1, 0){ .2000}}
\put( 0, 3.0000){\line( 1, 0){ .2000}}
\put(10, 3.0000){\line(-1, 0){ .2000}}
\put( 0, 4.0000){\line( 1, 0){ .2000}}
\put(10, 4.0000){\line(-1, 0){ .2000}}
\put( 0, 5.0000){\line( 1, 0){ .2000}}
\put(10, 5.0000){\line(-1, 0){ .2000}}
\put( 0, 6.0000){\line( 1, 0){ .2000}}
\put(10, 6.0000){\line(-1, 0){ .2000}}
\put( 0, 7.0000){\line( 1, 0){ .2000}}
\put(10, 7.0000){\line(-1, 0){ .2000}}
\put( 0, 8.0000){\line( 1, 0){ .2000}}
\put(10, 8.0000){\line(-1, 0){ .2000}}
\put( 0, 9.0000){\line( 1, 0){ .2000}}
\put(10, 9.0000){\line(-1, 0){ .2000}}
\put( 0,10.0000){\line( 1, 0){ .2000}}
\put(10,10.0000){\line(-1, 0){ .2000}}
\put( 0,11.0000){\line( 1, 0){ .2000}}
\put(10,11.0000){\line(-1, 0){ .2000}}
\put( 0,12.0000){\line( 1, 0){ .2000}}
\put(10,12.0000){\line(-1, 0){ .2000}}
\put( 0,13.0000){\line( 1, 0){ .2000}}
\put(10,13.0000){\line(-1, 0){ .2000}}
\put( 0,14.0000){\line( 1, 0){ .2000}}
\put(10,14.0000){\line(-1, 0){ .2000}}
\put( 0,15.0000){\line( 1, 0){ .2000}}
\put(10,15.0000){\line(-1, 0){ .2000}}
\put( 0, 5.0000){\line( 1,0){   .4000}}
\put(10, 5.0000){\line(-1,0){   .4000}}
\put( 0,10.0000){\line( 1,0){   .4000}}
\put(10,10.0000){\line(-1,0){   .4000}}
\put( 0,15.0000){\line( 1,0){   .4000}}
\put(10,15.0000){\line(-1,0){   .4000}}
\put( -1.9000, 4.9100){    .50}
\put( -1.9000, 9.9100){   1.00}
\put( -1.9000,14.9100){   1.50}
\put( 1.0000,15.0000){\line(1, 0){  .4000}}
\put( 1.4000,15.0000){\line(1, 0){  .4000}}
\put( 1.8000, 2.2986){\line(1, 0){  .4000}}
\put( 1.8000, 2.2986){\line(0,  1){12.7014}}
\put( 2.2000,  .2028){\line(1, 0){  .4000}}
\put( 2.2000,  .2028){\line(0,  1){ 2.0958}}
\put( 2.6000,  .2028){\line(1, 0){  .4000}}
\put( 3.0000,  .2028){\line(1, 0){  .4000}}
\put( 3.4000,  .3380){\line(1, 0){  .4000}}
\put( 3.4000,  .3380){\line(0, -1){  .1352}}
\put( 3.8000,  .6085){\line(1, 0){  .4000}}
\put( 3.8000,  .6085){\line(0, -1){  .2704}}
\put( 4.2000,  .4732){\line(1, 0){  .4000}}
\put( 4.2000,  .4732){\line(0,  1){  .1352}}
\put( 4.6000, 1.2845){\line(1, 0){  .4000}}
\put( 4.6000, 1.2845){\line(0, -1){  .8113}}
\put( 5.0000, 1.6902){\line(1, 0){  .4000}}
\put( 5.0000, 1.6902){\line(0, -1){  .4056}}
\put( 5.4000, 3.2451){\line(1, 0){  .4000}}
\put( 5.4000, 3.2451){\line(0, -1){ 1.5549}}
\put( 5.8000, 3.3803){\line(1, 0){  .4000}}
\put( 5.8000, 3.3803){\line(0, -1){  .1352}}
\put( 6.2000, 2.9747){\line(1, 0){  .4000}}
\put( 6.2000, 2.9747){\line(0,  1){  .4056}}
\put( 6.6000, 1.0141){\line(1, 0){  .4000}}
\put( 6.6000, 1.0141){\line(0,  1){ 1.9606}}
\put( 7.0000,  .1352){\line(1, 0){  .4000}}
\put( 7.0000,  .1352){\line(0,  1){  .8789}}
\put( 7.4000,  .0676){\line(1, 0){  .4000}}
\put( 7.4000,  .0676){\line(0,  1){  .0676}}
\multiput( 1.0000,15.0000)( .13333, 0.0){3} {\line(1,0){0.10}}
\multiput( 1.0000,15.0000)(0.0, .13333){    0}{\line(0, 1){0.10}}
\multiput( 1.4000,10.3538)( .13333, 0.0){3} {\line(1,0){0.10}}
\multiput( 1.4000,10.3538)(0.0, .13333){   34}{\line(0, 1){0.10}}
\multiput( 1.8000, 2.7147)( .13333, 0.0){3} {\line(1,0){0.10}}
\multiput( 1.8000, 2.7147)(0.0, .13333){   57}{\line(0, 1){0.10}}
\multiput( 2.2000,  .8207)( .13333, 0.0){3} {\line(1,0){0.10}}
\multiput( 2.2000,  .8207)(0.0, .13333){   14}{\line(0, 1){0.10}}
\multiput( 2.6000, 1.0101)( .13333, 0.0){3} {\line(1,0){0.10}}
\multiput( 2.6000, 1.0101)(0.0,-.13333){    1}{\line(0,-1){0.10}}
\multiput( 3.0000,  .6313)( .13333, 0.0){3} {\line(1,0){0.10}}
\multiput( 3.0000,  .6313)(0.0, .13333){    2}{\line(0, 1){0.10}}
\multiput( 3.4000, 1.4521)( .13333, 0.0){3} {\line(1,0){0.10}}
\multiput( 3.4000, 1.4521)(0.0,-.13333){    6}{\line(0,-1){0.10}}
\multiput( 3.8000, 1.4521)( .13333, 0.0){3} {\line(1,0){0.10}}
\multiput( 4.2000, 1.8940)( .13333, 0.0){3} {\line(1,0){0.10}}
\multiput( 4.2000, 1.8940)(0.0,-.13333){    3}{\line(0,-1){0.10}}
\multiput( 4.6000, 2.6516)( .13333, 0.0){3} {\line(1,0){0.10}}
\multiput( 4.6000, 2.6516)(0.0,-.13333){    5}{\line(0,-1){0.10}}
\multiput( 5.0000, 2.8410)( .13333, 0.0){3} {\line(1,0){0.10}}
\multiput( 5.0000, 2.8410)(0.0,-.13333){    1}{\line(0,-1){0.10}}
\multiput( 5.4000, 2.5253)( .13333, 0.0){3} {\line(1,0){0.10}}
\multiput( 5.4000, 2.5253)(0.0, .13333){    2}{\line(0, 1){0.10}}
\multiput( 5.8000, 1.3258)( .13333, 0.0){3} {\line(1,0){0.10}}
\multiput( 5.8000, 1.3258)(0.0, .13333){    8}{\line(0, 1){0.10}}
\multiput( 6.2000,  .3157)( .13333, 0.0){3} {\line(1,0){0.10}}
\multiput( 6.2000,  .3157)(0.0, .13333){    7}{\line(0, 1){0.10}}
\multiput( 6.6000,  .0631)( .13333, 0.0){3} {\line(1,0){0.10}}
\multiput( 6.6000,  .0631)(0.0, .13333){    1}{\line(0, 1){0.10}}
\put(0.,0.) {\line(1,0){10}}
\put(0.,15.) {\line(1,0){10}}
\put(0.,0.) {\line(0,1){15}}
\put(10.,0.) {\line(0,1){15}}
\put(4.8,-2.2) {$\Sigma(R)$}
\put(-2.9,12.0) {\Large $ \frac{N(\Sigma(R))}{V(\beta)}$}
\end{picture}
\end{center}
\vspace*{1.0cm}
\begin{center}
\parbox{14cm}{\caption{\label{histogram} Histograms
for summed Wilson action $\Sigma(\lambda^2,R$)
for $\beta$ = 2.85, lattice size = $48^3 \times 64$ (full line),
and for $\beta$ = 2.70, lattice size = $32^3 \times 48$ (broken line).
The summed Wilson action $\Sigma(\lambda^2,R)$ and $R$ are defined
in eq. \ref{sigma}. The histograms are normalized to V($\beta$),
the physical volume of the lattices.}}
\end{center}
\end{figure}

is shown as the solid line histogram of fig. \ref{histogram}.
The distribution is normalized by the physical volume $V(\beta)$
of the lattice (see below).
It is remarkable how clearly the peak at large
$\Sigma_m(\lambda^2,R)$ is separated from the background
at smaller values, with a dip around $\Sigma_m(\lambda^2,R) = 8. $
Furthermore, the data for case B are shown in fig. \ref{histogram} as
a broken line. In order to compare these data with those of case A,
one has to fix the ratios of the physical volumes. For this, I use
\begin{equation}
V(\beta = 2.7)/V(\beta = 2.85)
   = \frac{a(\beta =2.70)^4 \times 32^3 \times 48}{a(\beta = 2.85)^4
     \times 48^3 \times 64}
\end{equation}
with \cite{BOOTH} $ a(\beta =2.70) / a(\beta = 2.85) = 1.6$.

The fact that the dip is less pronounced at \be~=~2.7 than
at the higher value of \be, is, most likely, not due to
a statistical fluctuation.
Data at \be = 2.5 show that the peak for high values of
$\Sigma_m(\lambda^2,R)$ changes in shape towards a plateau.
At low \be, the spikes
are probably not as stable as at larger \be, and a snapshot
of a single configuration will find both well developed
spikes and decaying/growing ones.

It is natural to define the average
density of spikes per configuration by the number of objects
with  $\Sigma_m(\lambda^2,R) \geq 8$. In case A, this gives
\begin{equation}
\label{na}
N_{spikes}(\beta = 2.85) = 6.6 \pm 0.5
       \rm{\; spikes \;per\; configuration,}
\end{equation}
and, using the string tension scale \cite{BOOTH}
$a(\beta = 2.85) = 0.028 f\!m$,
\begin{equation}
\label{rho285}
\rho_{spikes}(\beta = 2.85) = 1.52\pm 0.2 \; {\rm spikes} \; / f\!m^4.
\end{equation}
The corresponding results for case B are:
\begin{equation}
N_{spikes}(\beta = 2.70) = 9.6 \pm 0.6 \rm{\;spikes\; per\; configuration }
\end{equation}
and
\begin{equation}
\rho_{spikes}(\beta = 2.70) = 1.51\pm 0.2 {\rm \;spikes}/f\!m^4.
\label{rho270}
\end{equation}

Thus, eqs. \ref{rho285} and \ref{rho270} show
with good accuracy, that the density of spikes
obeys scaling according to the
non-asymptotic variation of the string tension, and that the
density itself is a rather dilute one.
The spikes themselves are better defined for
large values of \be{} than for small ones, and this
implies, together with the scaling of the density, that they are not
a remnant of the strong coupling regime, which would become
irrelevant in the continuum limit.

It remains to show how the number of spikes varies when -at fixed \be- the
lattice size is chosen so small that one approaches
the deconfinement region
\footnote{The details of the transition to this
region depend on the lattice
geometry, and the present geometry is not optimally
chosen for this purpose.}.
I take the variation of certain glueball masses as a signal for
(de-)confinement,
especially the mass ratio of $A^+_1$ and $E^+$, which shows
a step\cite{TICKLE, VBAAL} as a function of the lattice size $L $ around
\begin{equation}
\label{za}
z _{A^+_1} = m _{A^+_1} L \approx 7.
\end{equation}
Combining this with the ratio of the root of the string tension,
$\sqrt{K},$
to the mass $m _{A^+_1}$, $\sqrt{K} / m _{A^+_1} \approx 0.25$, and
with the value of the string tension \cite{BOOTH}
$a^2 K(\beta = 2.85) = 0.0040$, I find for the deconfining lattice size
\begin{equation}
\label{Lstep}
L_{step}= 27.7 a.
\end{equation}
Now, the important question is how many spikes have to be expected on a
lattice of size $27.7^3\times 37$. As will be shown in the next section,
the density of spikes is (within an accuracy of 10\%)
independent of the volume.
Thus, from eq. \ref{na} we have to expect $0.73 \pm 0.15$ spikes per
configuration. This value allows to conclude
that at the deconfinement transition the average number of spikes
falls below unity.
\section{Gribov Copies and Spikes}
\label{gribov}
The existence of Gribov copies is made evident by applying
random gauge transformations prior to the Landau gauging.
Then, on large lattices, the final value of the gauge function
$F(U)$, eq. \ref{gauge}, will differ for each set of random numbers.
It remains to be seen how small the lattices have to be such that
the gauging procedure is practically unique in the sense
that for a large number of random trials one always ends up in
the same gauge. Here I will show that the naively estimated
number of spikes provides a good measure for the transition
from non-unique to unique gauging.

In detail, it will be studied how the appearance
of Gribov copies is correlated with the existence of spikes.
The study is restricted to a lattice size close to the
critical one for which the expected number of spikes
per configuration is just one,
in the following sense:
Given the number of spikes, $N_{spikes}$,
on a physically large lattice, $V_{large}$,
I define a critical volume by
\begin{equation}
V_{crit} = V_{large} / N_{spikes}.
\end{equation}
Thus, the number of spikes on $V_{crit}$  were one if
the density of spikes were independent of the volume,
which, of course, need not be the case a priori.

A lattice slightly smaller than the critical one seems
to be best suited to study the correlation between
the two phenomena. The reason is that
for lattices which are much larger than the critical one,
the number of Gribov copies cannot be determined reliably,
since for every random gauge one will end up in a different copy.
On the other hand, for lattices much smaller than the critical one,
copies are very rare, and it will take too many sweeps to find any.

At \be = 2.85, I found the average number of spikes on the large lattice,
eq.~\ref{na}, and accordingly we have
\begin{equation}
V_{crit} = (0.90 f\!m)^4.
\end{equation}
Thus, at \be{} = 2.85 and on a lattice of
size $24^3\times 32$, which corrsponds to a volume $V = (0.72 \fm)^4$,
we would expect 0.41 spikes by naive geometrical scaling.
The simulation, described in the following,
shows that the average number of spikes per configuration
is
\begin{equation}
\label{nsmall}
N_{spikes}(\beta = 2.85, 24^3\times 32)
=  0.37 \pm 0.03 \rm{\; spikes \;per\; configuration}.
\end{equation}
This is well compatible with the number $0.41$
expected from naive geometrical scaling. I conclude that -within
reasonable accuracy- the density of spikes is independent of the
lattice volume.

For the above lattice, the number of Gribov copies
for a given lattice configuration turns out
to be strongly correlated with the average number of spikes
observed on this configuration.
The observation is based on two long sequences
with about 200.000 updates for each sequence.
The approximate number of Gribov copies
has been determined after every 1.000 sweeps.
For this purpose, 12 random Landau gauges were performed,
and the 12 values of the gauge function were compared
and the numbers of different gauge functions,
found in this way, were recorded.
Simultaneously, the number of spikes was found by filtering
the configuration with a cut-off $\lambda^2 = 0.5a^2$
and by searching for local action maxima, which exceeded
the background by a factor three or more. For such maxima,
$\left|\hat{q}(x)\right| $ agrees with
the action within an accuracy of 10 \%.
The results for copies and spikes are the following:
\begin{itemize}
\item
Configurations without Gribov copies follow each other
in long sequencies, i.e. for $O(10.000)$ lattice updates. When spikes
occur, they preferentially show up in all random gauges, and very often
more than one spike is found. Thus, spikes and Gribov copies
occur, during the process of updating, in lumps.
\item
The appearance of Gribov copies is not exactly
correlated with the number of spikes, i.e. not on a 1:1 basis
for each configuration. There are configurations without
copies\footnote{Of course, it is possible that some copies might be found
if more random gauges had been tested.}, but with spikes, and vice versa.
\item
If both the number of copies and the number of spikes per configuration
are smeared over a few adjacent measured configurations (just
to improve the presentation), a striking correlation shows up.
This is demonstrated
in fig. \ref{histcorr}. On the left-hand side, the first 100.000 sweeps
are shown, with the slim curve giving the number of copies, and the fat
curve giving the number of spikes (averaged over the 12 random gauges).
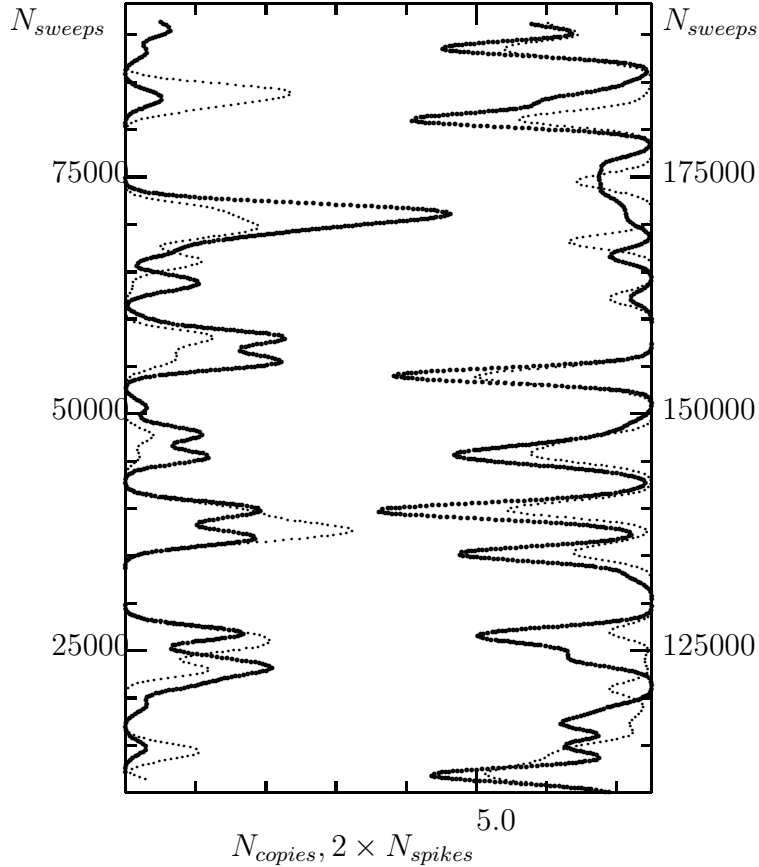
\begin{figure}[t]
\begin{center}
\unitlength0.7cm
\begin{picture}(10.,15.)
\normalsize
\thicklines
\input pic3.tex
\put(0.,0.) {\line(1,0){10}}
\put(0.,15.) {\line(1,0){10}}
\put(0.,0.) {\line(0,1){15}}
\put(10.,0.) {\line(0,1){15}}
\put(2.0,  -1.2)  {$\Large N_{copies}, 2 \times N_{spikes}$}
\put(-2.2, 14.5) {$N_{sweeps}$}
\put(10.2, 14.5) {$N_{sweeps}$}
\end{picture}
\end{center}
\vspace*{1.0cm}
\begin{center}
\parbox{14cm}{\caption{\label{histcorr} Averaged
number of Gribov copies per configuration (slim points)
and averaged number of spikes per configuration (fat points)
as function of number of sweeps in long updating sequence.
The lattice size is $24^3\times 32$, and \be = 2.85.
The Fourier cut-off is $\lambda^2 = 0.5 a^2.$}}
\end{center}
\end{figure}

The results from the next 100.000 sweeps are plotted
on the right hand side.
\end{itemize}

It is evident, that the appearance of Gribov copies
and of gauge singularities both do not stop
abruptly when the lattice becomes small.
Presumably, both phenomena are connected with the tunneling
of configurations from one "smooth" state to another one.

\section{Cooling a Gauged Configuration}
\label{cooling}
The nature of the spikes can be studied further if
smoothing by F.f. is confronted
with the results of the cooling technique.
By the latter, the quantum noise in gauge invariant
observables is reduced during the approach
to metastable minima of the action.
In this way, non-perturbative aspects of the
configuration, like instantons, may emerge.
Since instanton configurations
are close to local minima of the action, the distinction between
genuine instantons and artifacts of the cooling procdure,
is subjective. A comparison of the gauge fields etc.
of a cooled configuration with the same configuration modified by F.f.,
may be helpful to understand in what sense
instantons exist on the lattice.

For this purpose it is necessary to quantify the
impact of F.f. on instantons. This depends on the gauge.
For standard instantons in the regular gauge,
the effect of filtering is quite modest: The peak action of an
artificial instanton of radius $7 a$ (this corresponds to a
radius $\rho = 0.20 \fm $ at \be = 2.85) will be reduced by F.f.
not too strongly, if the filtering is done with a strong cut-off,
i.e. $\lambda^2 = 4.0 a^2$. The reduction amounts to a factor
\begin{equation}
\label{peakred}
S(x = 0,\lambda^2 = 4.0a^2) / S(x = 0,\lambda^2 = 0.50a^2) \approx 0.49.
\end{equation}
Thus, F.f. will not seriously
affect those instantons, which are the most numerous ones
in SU(2), according to ref. \cite{DEGRAND}.
Smaller ones will feel a stronger reduction in peak height, such that
they essentially look like broader instantons.

On the other hand, as has been stated already,
instantons in the singular gauge will show up, under F.f.,
as spikes and, therefore, cannot be missed. I assume, more generally,
that F.f. will not reduce the number of visible instantons,
if the creation and observation of spikes is properly taken into account.
Furthermore, F.f. will not create quasi-instantons nor improve the
self-dual properties of maxima, in obvious contrast to the cooling technique.

I have studied correlations between spikes and cooled objects
in two ways. First of all, around the location
of spikes in a given filtered configuration
(denoted by $\cal{F}$), the cooled configuration is studied visually.
Also, action maxima are studied, which are not associated with spikes.
Secondly, the inverse correlation will be investigated,
i.e. maxima in the cooled configuration, called  $\cal{C}$,
are selected and the filtered configuration $\cal{F}$ is visualized
around the corresponding locations. In the final subsection,
the probability for obtaining spikes by gauging cooled
configurations is investigated. The number of spikes and -to a
large degree- their positions seem to be independent
on the amount of cooling.
\subsection{Spikes $\Rightarrow$ Cooled Configuration}
The cooling sequence starts from the
same Landau gauged configurations as the filtered ones.
Four configurations have been cooled both with 30 and 100 steps,
using "strong" cooling. There, each step rotates each link variable
to the local action minimum.
A total of 37 spikes has been found, out of an ensemble of
128 action maxima.

For the spikes, the resulting correlation
between the filtered configuration and
the cooled one is quite simple:
At the location of the 36 spikes one finds,
first of all, always gauge singularities in $\cal{C}$,
and, secondly, one observes that the spikes are, apart from
7 cases\footnote{An inspection of these
exceptions reveals that either the cooled maximum of \qs{}
is quite low, or that cooling leads to an annihilation of a
instanton-anti-instanton pair.}, associated with instantons.

The first observation means that
the cooled gauge fields $A^a_{\mu}(x)$
show clean peaks around the position of the spikes,
including sign flips in all colours and directions,
with a diameter of 3 or 4 lattice unit.
Outside of the peaks, the gauge fields are noisy
and not particularly small.
This is due to the fact that cooling does not minimize
the gauge fields, and Landau gauging still preserves
their perturbative fluctuations.
The positions of the gauge singularity
and the nature of the gauge field peaks are
independent of the number of cooling steps,
as has been checked by comparing the fields
for 30 and 100 cooling steps.

The second observation means that at the position of a spike,
pronounced maxima of the action and of the topological charge density
show up in $\cal{C}$. The topological charge density has the opposite
sign in the two cases cooling and filtering\footnote{This
is in agreement with the interpretation
of spikes as a mismatch of the quadratic terms and the higher
order terms, caused by suppression of high frequency terms via filtering.}.
After cooling, the hights of the action-peaks
vary strongly from peak to peak. The explanation is that for instantons,
the height of the action-maxima is a strongly
decreasing function of the instanton size, such that large instantons
do not show up spectacularly under visualization techniques. Obviously,
selecting positions by spikes does not select instantons of a
particular radius.

For action-maxima observed after F.f. which are not spikes, the situation
at the same position in $\cal{C}$ is not very clear-cut.
At some maxima one may find an instanton-like object,
at others there is one close by,
and in many cases one cannot observe any activity in a
neighbourhood of reasonable size.

\subsection{Maxima of Cooled Configuration $\Rightarrow$ Filtered
Configuration.}
The inverse correlation, i.e. between cooled
action (or \qsa) maxima
selected in $\cal{C}$ on the one hand,
and between structures in $\cal{F}$ (obtained by filtering)
on the other hand, is more complicated
to investigate than the previous one.
This holds because there is much freedom in the selection of maxima.
It is not the purpose of this paper to follow the
elaborate filtering techniques of ref. \cite{TEPER} for
extracting the best instanton candidates among the many
maxima which show up during the cooling process.
I simply start from the 4 configurations
which have been "strongly" cooled
by 100 sweeps, select for each configuration 32 positions
associated with the 32 highest maxima of \qsa,
and investigate the properties of $\cal{F}$
around the positions of these maxima. The filtering is done
both with parameters $\lambda^2 = 4a^2$ and $\lambda^2 = 0.5a^2$.

The first observation is that one recovers $\approx 80\%$ of the
spikes among the first $O(30)$ maxima.
This is not trivial. The sizes of the instantons which are associated
with spikes, vary considerably, such that a selection via the
\qsa-peak height might lead to a
failure in identifying these maxima within the first few dozens of peaks.
I conclude that the spikes are tightly correlated
with instantons which are produced by cooling and selected
among the spacially less extended ones.
Obviously, the spikes have a significance beyond the gauge dependent
filtering technique.

Next, it is highly interesting to investigate those locations
in the filtered configuration, $\cal{F}$, which on the one hand
are associated with peaks in $\cal{C}$ but, on the other hand,
have no singularity in the gauge fields at this location.
In a fraction of about 2/3 of those maxima,
one encounters also a peak in $\cal{F}$ both in the action and in \qsa,
with identical signs of \qs{} in both configurations,
i.e. (anti-)instantons in non-singular
gauge in $\cal{C}$ may be associated with candidates for
(anti-)instantons in $\cal{F}$.
In the other 1/ 3 cases, there is no significant
activity in $\cal{F}$.

Now, the crucial question is whether these candidates
have the correct properties to be unambiguously
identified with instantons. This has to be doubted
for the following reasons:
\begin{itemize}
\item
In $\cal{F}$, the orientation of the colour components
in colour space is not along the diagonal. This fact is best
recognized by visual inspection,
and it will not be specified in detail here.
\item
The electric and magnetic fields in $\cal{F}$
are not perfectly self-dual.
In order to be quantitative, I consider the
measure of self-dual quality, $Q^{(a)}(y)$, defined by
\begin{equation}
Q^{(a)}(y) = \frac{2\vec{E^{(a)}}(y) \vec{B^{(a)}}(y)}
{\vec{E^{(a)}}^2(y) + \vec{B^{(a)}}^2(y)}, \;\;\; a = 1,2,3.
\end{equation}
where $y$ is taken at the lattice positions
around the maxima position $x$ in $\cal{C}$,
with $\left|x-y\right|~<~4a$.
This distance amounts to half the radius of a typical SU(2)-instanton.
A histogram of this quantity, taken between  $ -1 \geq Q^{(a)}(y)\geq 1$,
is  peaked beautifully at the limits $\pm1$ for the {\bf cooled}
configuration. For the {\bf filtered} configuration, however, $Q^{(a)}(y)$
shows a broad histogram for all $ a $,
extending down to $Q^{(a)}(y)\approx 0$.
Thus, self-duality is not realized well in the case of F.f..
\item
The spatial sizes of the peaks in $\cal{F}$
are much smaller than in $\cal{C}$. Since the shape of the topological
objects seems to be rather irregular, their radii are difficult to
determine directly. A simpler way is to observe the reduction of the
peak height as a function of the filtering parameter, when it
is increased from $\lambda^2 = 0.5a^2$ to
$\lambda^2 = 4a^2$. This reduction amounts approximately
to a factor 0.03, in sharp contrast to the reduction by
a factor 0.49 for an artificial lattice instanton (see eq.~\ref{peakred}).
\end{itemize}

It has to be concluded that a close correspondence
between cooled configurations and filtered ones
exists mainly at the position of the spikes.
In the cooled configuration, the gauge singularity
-the origin of the spike- is preserved,
and an instanton-like maximum of the action etc. has developed.
Action-maxima in $\cal{C}$, which are not associated with spikes,
and which may therefore be displayed in the filtered configuration,
do not reveal the typical properties of instantons.
Of course, by a more refined search among the many maxima
in $\cal{C}$ one may eventually find better instanton candidates.

\subsection{Gauge and Cooling (In-)Dependence of Spike Positions}
In the previous subsection, it has been stated that
almost all spikes in the filtered configurations
are found at a position close to maxima
in the cooled configurations. The relevant maxima are those
with the largest values of the action density.
This is rather important, since the positions of these
cooled maxima are gauge independent. Thus, also the position
of spikes has a gauge invariant meaning, at least in
some probabilistic sense.

Furthermore, in the process of cooling,
most small scale fluctuations are eliminated which,
in principle, could induce the gauge singularities.
The presence of such "dislocations" which can contribute
to the topological charge \cite{FORCRAND, GOCKEL}
is a drawback of using the Wilson action both for
updating configurations and for cooling.

In the following,
the effect of cooling will be studied once more in a different way.
In determining the correlation between the cooled maxima and the
spikes, the latter were defined by Landau gauging a configuration
which had all short range fluctuations undamped.
Here, the order of cooling and gauging will be reversed,
i.e. a non-gauged configuration will be cooled
with up to 10 strong cooling steps. This eliminates
all plaquette values smaller than 0.9.
If one then finds a Landau gauge and applies filtering
to this cooled configuration, one finds approximately
the same number of spikes as compared to the
case without cooling. Most of the spikes show up at the
same positions in the two cases.

In detail, on a lattice of
size $24^3 \times 32$ at $\beta = 2.65$, 20 configurations
have been cooled with 3 steps (case (a)), and 20 configurations with
10 steps (case (b)). After a random gauge had been applied,
first the Landau gauging and then filtering were performed. In case (a)
one finds 93 spikes, whereas for the corresponding uncooled
configurations one finds 87 spikes. In case (b),
the corresponding numbers are 99 and 91.
When the spike positions are compared between the cooled
and the uncooled configurations, it turns out that in 15 configurations
out of the 20 cooled ones (in case (b)), more than 50\% of the
spikes show up at the same positions as in the uncooled case
(within 1 or 2 lattice units).

Because of the strong suppression of plaquettes with trace values
$< 0.9$, this correlation of spike positions makes it rather
unlikely that the spikes are induced by dislocations,
in so far as these are characterized
by large negative plaquette values.

\section{The Gauge Field Propagator and Spikes}
\label{correlations}
It is evident that close to a spike, the gauge fields show a rapid flip of
sign, for all colours, for all space time orientations
and along all directions.
Thus, it is natural to expect that the gauge field correlators
behave differently for configurations with spikes as compared
to configurations without spikes. A convenient tool to
study this effect is a measurement of the gluon propagator,
evaluated separately for the two specimens of configurations.
It is to be expected that the propagator in x-space falls off
more rapidely for increasing spatial separation, when spikes are present
than if none are around. In momentum space, this effect is reproduced
if the zero momentum propagator is reduced and the
small -but non-zero- momentum propagator
is enhanced when spikes are present,
as compared to the spike-free configurations.

This can be tested on lattices
which are so small that -simply by geometrical considerations-
the probability to find a spike is considerably smaller than one.
For our standard value of \be = 2.85, a lattice of size $24^3\times 32$
has a density of 0.4 spikes per configuration (see eq. \ref{nsmall}).
Since often there are more
than one spike per configuration, the probability to find at least
one spike is around 0.2, i.e. considerably smaller than 0.4. Thus for a
total of 600 configurations which have been
measured\footnote{Out of 300.000 sweeps,
after 500 updates Landau gauging and searching
for spikes was performed.}, one finds 480 configurations
without spikes and 120 configurations with one or more spikes.
The gauge field propagator has been measured separately for the two classes
of configurations.

The difference between the two values of the propagator,
-spikes present or not- depends on the momentum $q^2$. For $q^2 = 0$,
the propagator with spikes is slightly smaller than without spikes:
\begin{equation}
\label{diffq=0}
G(q^2=0)_{no\;spikes} \approx (1.19 \pm 0.05 ) G(q^2=0)_{spikes}.
\end{equation}
This difference is just significant.
For $ q^2 > 0$, the error is significantly smaller, and we have
\begin{equation}
\label{diffq>0}
G(q^2 \approx 0.1 a^{-2})_{no\,spikes} \approx (0.78 \pm 0.03 )
G(q^2 \approx 0.1 a^{-2})_{spikes}.
\end{equation}

For $q^2 > 1/a^2$, no significant difference between the
propagators of the two classes could be observed.
It is straightforward to transform the
different behaviour of the propagators
to gauge field correlators in x-space. If one sums the gauge fields
over 3 spatial dimensions and considers the correlator of this average
along the t-direction,
\begin{equation}
C(T) = <\sum_{\vec{x}}A(t, \vec{x})
\sum_{\vec{x}}A(t+T, \vec{x})>/<(\sum_{\vec{x}}A(t, \vec{x}))^2>,
\end{equation}
then the inequalities \ref{diffq=0} and \ref{diffq>0}
imply that the correlator with spikes decreases faster with T that
the correlator without spikes. The effect is small but significant.
\section{Conclusions}
\label{summary}
Configurations of SU(2) lattice theory, when transformed to a Landau gauge,
reveal special points with a scale invariant \footnote{This holds if the
scale as a function of \be{} is taken
from a measurement of the string tension.}
density of 1.5 points per $f\!m^4$, where the
gauge fields show a singularity, regulated by the lattice.
This singularity clearly shows up either
if the gauged configuration is cooled -with $O(50)$
cooling steps- or if the high momentum Fourier components are
suppressed by some exponential cut-off.
In detail, the phenomena are:
\begin{itemize}
\item
After cooling a gauged configuration, the gauge fields
resemble those of a singular gauge (see eq. \ref{singau}),
i.e. they shoot up and change sign at the special points.
The gauge invariant action and the topological
charge density signal the appearance of instantons of various sizes
around those singularities, with the gauge fields in the singular gauge.
Other -nonsingular- instanton-like objects show up
visually with a density which is higer than that of the singular ones.
This density drops quickly under prolonged cooling.
The instantons associated with singularities
are stable under cooling.
\item
After removing the high momentum amplitudes by
Fourier filtering, the singularities show up as spikes in the Wilson
action and in the topological charge density.
This is so because the removal
of short range Fourier amplitudes destroys the cancellation between
linear and quadratic terms in the field strengths. The gauge fields
show zeroes, as a function of lattice positions,
in all colours and directions, but the peaks are smoothed out
relative to the cooling procedure. When more and more Fourier
amplitudes are removed, the spikes get, of course, broader.
However, the positions do not vary essentially.
\item
The positions of the spikes
are not completely independent of the special Landau gauge,
but almost so. This means that for different Gribov copies
of a given configuration, almost all spikes appear at the same position,
with a mismatch in the order of $10\%$ (see \cite{GUTBROD1}). Furthermore,
the spikes are strongly correlated with the -gauge invariant- positions
of the leading action maxima which are generated by cooling.
\item
The density of spikes is correlated with the appearance of
of Gribov copies. This holds in the sense
that on physically small lattices,
where Landau gauging is almost unique and where the probability
for the occurrence of spikes is smaller than unity,
Gribov copies preferentially show up in the
same configurations as spikes do.
\item
The gauge field propagator is different for configurations with spikes
as compared to the case without spikes. This shows up as a faster
temporal decrease of zero momentum gauge field correlators.
\end{itemize}

In summary, it is evident that the presence of spikes is strongly
correlated with other nonperturbative phenomena on the lattice.
In particular, the correlation with the gluon propagator implies that
the presence of spikes is connected with the decrease of the
propagator in x-space. The sign flip of gauge fields, which
are associated with gauge singularities, intuitively provides
a mechanism for a fast decorrelation of the fields. According to
ref. \cite{DOSCH}, such a decorrelation can be responsible for
deconfinement. A study of the correlation of the string tension
with the spike density on larger lattices seems to be topic for
future investigations.
\vspace{0.5cm}\\
{\large \bf Acknowledgement}
\normalsize
\vspace{0.1cm}\\
The author is indebted to H. Joos, I. Montvay,
G. Schierholz, R.L. Stuller and H. Wittig for useful
discussions and for encouragement.
The preparation of the SU(2)-configurations on the large lattice
has been accomplished on 108 nodes of the T3E parallel computer at the NIC
at the Forschungszentrum J\"ulich. The author is grateful for granted
computer time and support, especially to H. Attig.
The development of the visualization tools has been carried out on a
dedicated SNI Celsius workstation.
The author is indebted to the DESY Directorium for providing
access to this facility.

\end{document}